\documentclass[aps,prc,twocolumn,amsmath,amssymb,showpacs]{revtex4}
\usepackage{graphicx}
\usepackage{mathptmx}
\usepackage{epsfig}
\usepackage{dcolumn}
\usepackage[T1]{fontenc}

\begin{document}

\title{Extraction of thermal and electromagnetic properties in $^{45}$Ti}

\author{N.U.H.~Syed$^1$\footnote{Electronic address: n.u.h.syed@fys.uio.no}, A.C.~Larsen$^1$, A.~B\"{u}rger$^1$, M.~Guttormsen$^1$, S.~Harissopulos$^2$, M.~Kmiecik$^4$, T.~Konstantinopoulos$^2$, M.~Krti\v{c}ka$^3$, A.~Lagoyannis$^2$, T.~L\"{o}nnroth$^5$, K.~Mazurek$^4$, M.~Norby$^5$, H.~T.~Nyhus$^1$, G.~Perdikakis$^2$\footnote{Current address: National Superconducting Cyclotron Laboratory, Michigan State University, East Lansing, MI 48824-1321, USA.}, S.~Siem$^1$, and A.~Spyrou$^2$\footnotemark[\value{footnote}]\\}

\affiliation{$^1$Department of Physics, University of Oslo, P.O.Box 1048 Blindern, N-0316 Oslo, Norway.\\
$^2$Institute of Nuclear Physics, NCSR "Demokritos", 153.10 Aghia Paraskevi, Athens, Greece.\\ 
$^3$ Institute of Particle and Nuclear Physics, Charles University, Prague, Czech Republic.\\
$^4$ Institute of Nuclear Physics PAN, Krak\'{o}w, Poland.\\
$^5$Department of Physics, \AA bo Akademi, FIN-20500 \AA bo, Finland.\\}

\date{\today}

\begin{abstract}
The level density and $\gamma$-ray strength function of $^{45}$Ti have been determined by use of the Oslo method. The particle-$\gamma$ coincidences from the $^{46}$Ti($p,d\gamma$)$^{45}$Ti pick-up reaction with 32 MeV protons are utilized to obtain $\gamma$-ray spectra as function of excitation energy. The extracted level density and strength function are compared with models, which are found to describe these quantities satisfactorily. The data do not reveal any single-particle energy gaps of the underlying doubly magic $^{40}$Ca core, probably due to the strong quadruple deformation.
\end{abstract} 



\maketitle

\section{Introduction}
The density of nuclear levels in quasi-continuum provides information on the gross structure of excited nuclei and is a basic quantity in nuclear reaction theories. Also the $\gamma$-ray strength function is important for describing the $\gamma$-decay process at high excitation energies. Experimentally, the nuclear level density can be determined reliably up to a few MeV of excitation energy from the counting of low-lying discrete known levels~\cite{ENSDF}. Previously, the experimental information on the $\gamma$-ray strength function has been mainly obtained from the study of photonuclear cross-sections~\cite{dietrich}.

The Oslo nuclear physics group has developed a method to determine simultaneously level densities and $\gamma$-ray strength functions from particle-$\gamma$ coincidences. The "Oslo method", which is applicable for excitation energies below the particle separation threshold, is described in detail in Ref.~\cite{schl0}. In this work, we report for the first time on results obtained using the $(p,d)$ reaction as input for the Oslo method. The advantage of this reaction compared to the commonly used ($^3$He,$^3$He') and ($^3$He,$^4$He) reactions, is a higher cross-section and a better particle energy resolution.

The subject of this paper is to present the level density and average electromagnetic properties of the $^{45}$Ti nucleus. The system has only two protons and three neutrons outside the doubly magic $^{40}$Ca core. It is therefore of great interest to see if the number of levels per MeV is quenched due to the low number of interplaying valence nucleons. Also the decay pattern may be influenced by the expected overrepresentation of negative parity states, originating from the $\pi f_{7/2}$ and $\nu f_{7/2}$ single-particle states.

In Sect.~II, the experimental set-up and data analysis are described. Nuclear level densities and thermodynamics are discussed in Sect.~III, and in Sect.~IV, the $\gamma$-ray strength function is compared with models and photonuclear cross-section data. Summary and conclusions are given in Sect.~V.

\section{Experimental setup and data analysis}

The experiment was conducted at the Oslo Cyclotron Laboratory (OCL) using a 32-MeV proton beam impinging on a self-supporting target of $^{46}$Ti. The target was enriched to 86$\%$ $^{46}$Ti (10.6$\%$ $^{48}$Ti, 1.6$\%$ $^{47}$Ti, 1.0$\%$ $^{50}$Ti, and 0.8$\%$ $^{44}$Ti) and had a thickness of 1.8 mg/cm$^2$. The transfer reaction, $^{46}$Ti(p,d$\gamma$)$^{45}$Ti, is analyzed in the present study. The deuterium ejectile is used to identify the reaction channel.

Particle-$\gamma$ coincidences were measured with the CACTUS multi-detector array~\cite{Cactus}. The coincidence set-up consisted of eight collimated $\Delta$E -- E type Si particle telescopes, placed at 5 cm from the target and making an angle of 45$^{\circ}$ with the beam line. The particle telescopes were surrounded by 28 $5^"\!\times 5^"$  NaI $\gamma$-ray detectors, which have a total efficiency of $\sim$ 15.2$\%$ for the 1332-keV $\gamma$-ray transitions in $^{60}$Co.

The experimental extraction procedure and the assumptions made are described in Ref.~\cite{schl0}.  The registered deuterium ejectile energy is transformed into the initial excitation energy of the residual nucleus through reaction kinematics and the known reaction $Q$-value. The excited residual nucleus produced in the reaction will subsequently decay by emission of one or more $\gamma$-rays, provided that the excitation energy is not much above the particle separation energy. Thus, a $\gamma$-ray spectrum can be recorded for each initial excitation energy bin $E$. In the analysis, the $\gamma$-ray spectra are corrected for the NaI detector response function by applying the unfolding technique of Ref.~\cite{gutt6}. The unfolding uses the Compton-subtraction method, which prevents additional count fluctuations from appearing in the unfolded spectrum. 

The set of these unfolded $\gamma$-ray spectra, which are organized into a $(E, E_\gamma)$ matrix, forms the basis for the extraction of the first-generation $\gamma$-ray spectra. Here, the energy distribution of the first (primary) emitted $\gamma$-rays in the $\gamma$-cascades at various excitation energies is isolated by an iterative subtraction technique~\cite{gutt0}. The main assumption made for the first-generation method is that the $\gamma$-ray spectrum from a bin of excited states is independent of the population mechanism of these states. In the present setting, this means that the $\gamma$-spectrum obtained from a direct $(p,d)$ reaction into states at excitation energy bin $E$ is similar to the one obtained if the states at $E$ are populated from the $\gamma$-decay of higher-lying states. 

According to the generalized Fermi's golden rule the decay probability can be factorized into a function depending on the transition matrix-element between the initial and final state, and the state density at the final states. Following this factorization, we can express the normalized primary $\gamma$-ray matrix $P(E,E_\gamma)$, which describes the $\gamma$-decay probability from the initial excitation energy $E$, as the product of the $\gamma$-ray transmission coefficient $\cal {T}$$(E_\gamma)$ and the level density $\rho(E-E_\gamma)$,
\begin{equation}
P(E, E_{\gamma}) \propto   {\cal{T}}  (E_{\gamma})  \rho (E -E_{\gamma}).\
\label{eqn:2}
\end{equation}
Here, ${\cal T}(E_\gamma)$ is assumed to be temperature (or excitation energy) independent according to the Brink-Axel hypothesis~\cite{brink,axel}. 

The functions $\rho$ and $\cal {T}$ are determined in an iterative procedure~\cite{schl0} by adjusting the functions until a global $\chi^2$ minimum with the experimental $P(E, E_{\gamma})$ matrix is reached. It has been shown~\cite{schl0} that if one of the solutions for $\rho$ and $\cal {T}$ is known, then the entries of the matrix $P(E, E_\gamma)$ in Eq.~(\ref{eqn:2}) are invariant under the transformations:
\begin{equation}
\tilde{\rho}(E-E_\gamma) = A\exp[\alpha(E-E_\gamma)]\rho(E-E_\gamma),\
\label{eqn:4}
\end{equation}
\begin{equation}
\tilde{\cal T}(E_\gamma) = B\exp(\alpha \ E_\gamma) \ {\cal {T}} (E_\gamma).\
\label{eqn:5}
\end{equation}
The parameters $A$, $B$ and $\alpha$ are unknown and can be obtained by normalizing our data to other experimental data or systematics. The determination of $A$ and $\alpha$ is discussed in the next section, and the parameter $B$ is discussed in Sect.~IV.

\section{Nuclear level density and thermodynamics}

The level density $\rho$ extracted from our coincidence data by use of Eq.~(\ref{eqn:2}) represents only its functional form and is not normalized. The transformation generators $A$ and $\alpha$ of Eq.~(\ref{eqn:4}), which give the absolute value and slope of the level density, can be determined by normalizing $\rho(E)$ to the discrete levels at low excitation energies~\cite{ENSDF} and to the level density determined from available resonance spacing data of nucleon capture experiments. Unfortunately, the neutron-resonance spacing data for the target nucleus $^{44}$Ti are not found in literature ($^{44}$Ti has a half life of 67 years). We therefore use the systematics of T.~von Egidy and D.~Bucurescu~\cite{BSFG}, which is based on a global fitting of known neutron resonance spacing data with the back-shifted Fermi gas (BSFG) formula:
\begin{equation}
\rho_{\rm BSFG}(U) = \frac{\sqrt{\pi}}{12}\frac{\exp(2\sqrt{aU})}{a^{1/4}U^{5/4}}\frac{1}{\sqrt{2\pi}\sigma},
\label{eqn:6}
\end{equation}
where $a$ is the level density parameter and $U=E-E_1$ is the intrinsic excitation energy, $E_1$ being the back-shifted energy parameter. The spin distribution is given by the spin cut-off parameter $\sigma$ with
\begin{equation}
\sigma^2 = 0.0146 A^{5/3} \frac {1 + \sqrt{1+ 4 a (E-E_1)}} {4a},
\label{eqn:8}
\end{equation}
where $A$ is the nuclear mass number. The level-density parameters are summarized in Table~\ref{tab:tab1}. 

\begin{figure}
\centering
\includegraphics[height=0.8\textheight]{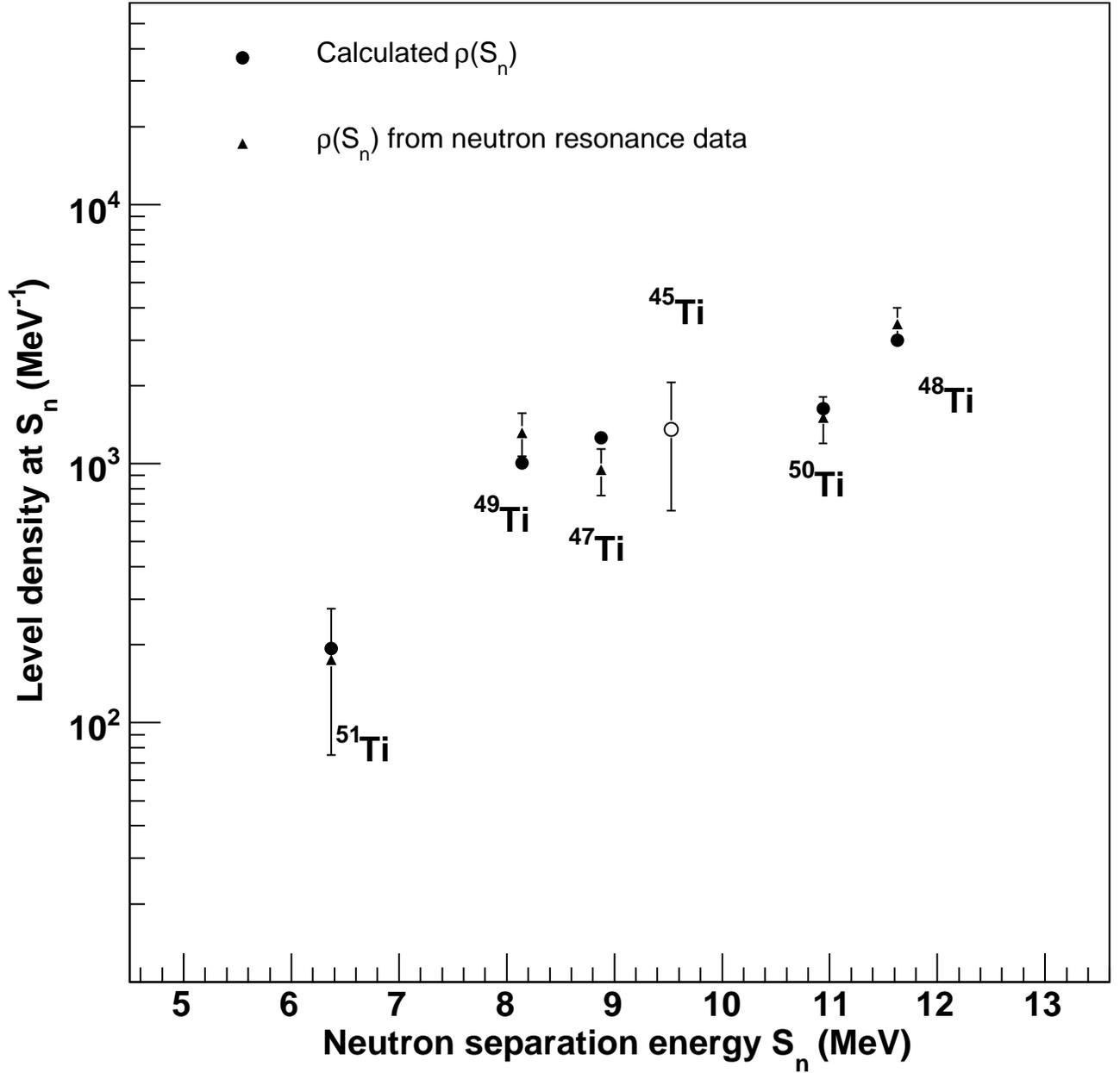}
\caption{Neutron resonance spacings data of~\cite{RIPL} have been used to deduce $\rho(S_n)$ (filled triangles). The filled circles represent the level densities that have been calculated using Eq.~(\ref{eqn:6}). The calculated level densities are multiplied with a factor of 0.5 in order to improve the agreement with level densities deduced from resonance spacing data. The level density of $^{45}$Ti (open circle) is calculated and scaled using the same systematics, assuming an uncertainty of 50$\%$.}
\label{fig:estimation}
\end {figure}

\begin{figure}
\centering
\includegraphics[height=0.8\textheight]{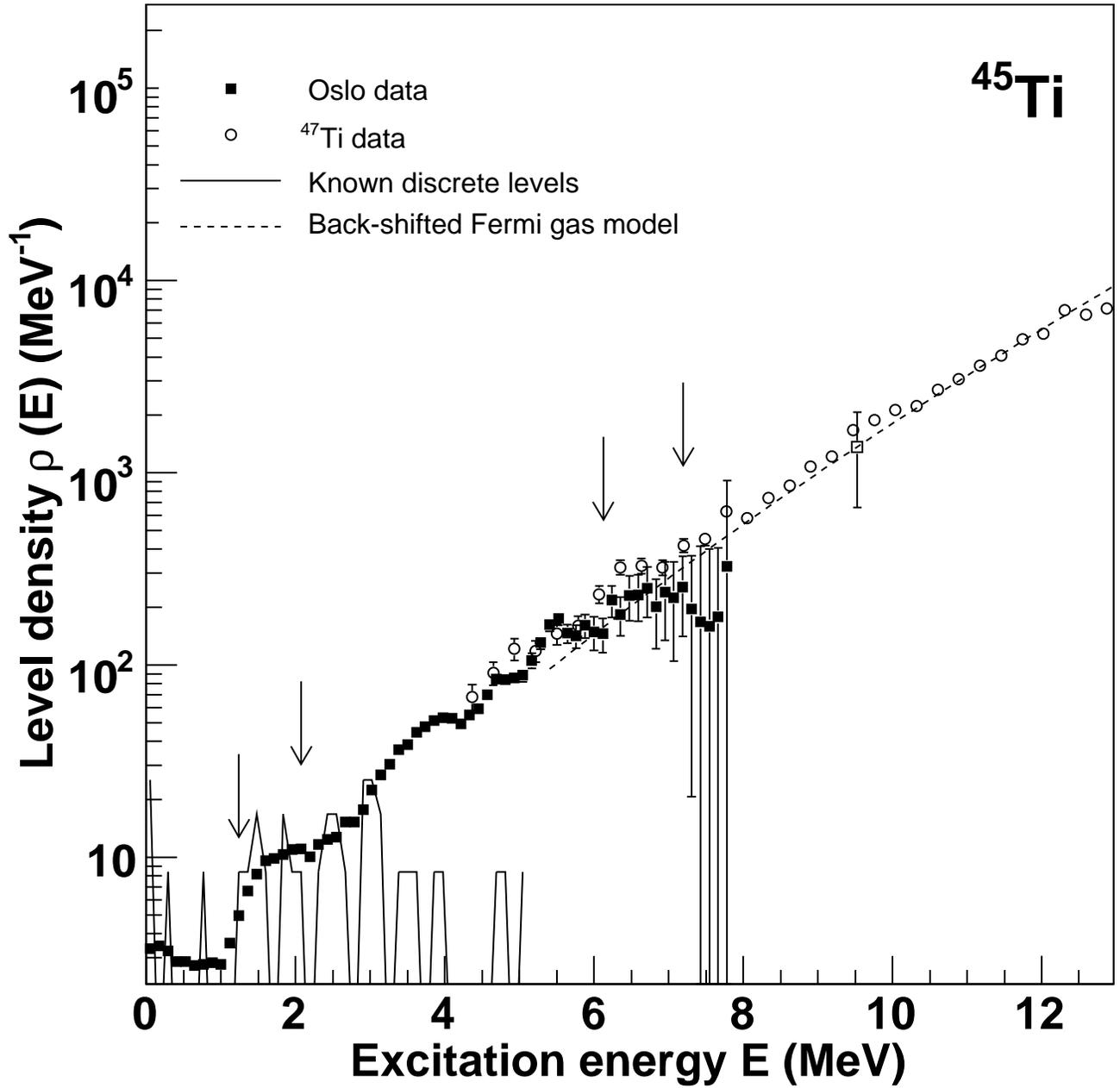}
\caption{Normalization of the nuclear level density (filled squares) of $^{45}$Ti. At low excitation energies the data are normalized (between the arrows) to known discrete levels (solid line). At higher excitation energies, the data are normalized to the BSFG level density (dotted line). The open square is the level density at $S_n$ estimated from the systematics of Fig.~\ref{fig:estimation}. Open circles represent data from particle-evaporation spectra~\cite{voinov2}.}
\label{fig:nld1}
\end {figure}

In Fig.~\ref{fig:estimation} nuclear level densities deduced from the neutron resonance spacing data~\cite{RIPL} of odd and even titanium nuclei are compared with BSFG level densities obtained from systematics~\cite{BSFG}. The systematic values $\rho(S_n)$ for the titanium isotopes overestimate the experimental values determined from resonance spacing data~\cite{RIPL}. Thus, the systematic values from Eq.~(\ref{eqn:6}) are multiplied with a factor of $\eta=0.5$ to improve the agreement with the experimental values. In this way (see Fig.~\ref{fig:estimation}), the unknown level density of $^{45}$Ti can be estimated as $\rho(S_n) \sim 1400 \pm 700$ MeV$^{-1}$. The high uncertainty for the deduced $\rho(S_n)$ in $^{45}$Ti is chosen so that it corresponds to the combined deviation of the systematics from the experimental data. Figure~\ref{fig:nld1} shows our normalized level density for $^{45}$Ti, which is adjusted in $A$ and $\alpha$ to fit the discrete levels at low excitation energy and the estimated $\rho(S_n)$.

The slope (given by $\alpha$) of the measured level density has a large uncertainty because of the estimated level density at $S_n$, the upper anchor point of the normalization. Various ways of analyzing the global set of neutron resonance spacing data may result in $\rho(S_n)$ values differing by a factor of two. Additionally, the uncertainty of the slope of the normalized level density is also affected if one chooses different excitation energy intervals during normalization. However, we have chosen an interval as high in excitation energy as possible with respect to reasonable errors in the data points. By choosing an interval 1 MeV lower, a decrease of $\sim$35\% in $\rho$ around 7 MeV is obtained. The normalization around 2 MeV of excitation energy is settled by known levels. It has been observed in many analyzed nuclei like Sc, V, Fe, and Mo~\cite{Sc,V,Fe,Mo} that level densities of neighboring isotopes show a comparable excitation energy dependence. So, in order to be more confident with our level-density normalization, the experimental data of $^{47}$Ti~\cite{voinov2} are compared with our data of $^{45}$Ti, as shown in Fig.~\ref{fig:nld1}. The level density in $^{47}$Ti is determined using proton-evaporation spectra from the $^{45}$Sc($^3$He,p)$^{47}$Ti reaction. The good agreement between the slopes for $^{45}$Ti and $^{47}$Ti in the energy region of 4 -- 8 MeV gives confidence to the present normalization.

The level density obtained from the present experiment agrees well with the density of known discrete levels up to $E \sim 2-3$ MeV. At higher excitation energies, only a part of the levels are known, mainly from transfer reactions. Our experimental data points only reach up to an excitation energy of $E \sim S_n -1$ MeV due to methodological difficulties regarding the first generation $\gamma$-ray extraction procedure. The gap between our data points and $\rho(S_n)$ is interpolated by the BSFG model
\begin{equation}
\rho_{BSFG}(E) = \eta \frac{\exp(2\sqrt{aU})}{12 \sqrt{2}a^{1/4}U^{5/4}\sigma}, \ 
\label{eqn:BSFG}
\end{equation}
where the constant $\eta$ ensures that the BSFG level density crosses the estimated $\rho(S_n)$ at the neutron separation energy. As shown in Fig.~\ref{fig:nld1} the BSFG level density in the extrapolated region only describes the general increase in level density and not the fine structures as described by our data.

The level density of $^{45}$Ti shows pronounced step structures below $E \sim 6-7$ MeV. These structures probably correspond to the amount of energy needed to cross shell gaps and to break Cooper pairs~\cite{BCS}. The gap energies are around 2 -- 3 MeV, which is similar to the energy ($2\Delta$) required for breaking Cooper pairs. Thus it is very difficult to foresee the level-density fine structure in this case.

From the measurements of level density as a function of excitation energy, one can explore thermodynamic properties of the nucleus. The starting point is the entropy of the system, from which one may deduce temperature, heat capacity and single particle entropy. The multiplicity of states $\Omega_s$, the number of physically accessible microstates, is related to the level density and average spin $\langle J(E) \rangle$ by
\begin{equation}
\Omega_s(E) \propto \rho(E) \left[ 2 \langle J(E) \rangle +1 \right].
\label{eqn:5.1}
\end{equation} 
The $2J+1$ degeneracy of magnetic sub-states is not included, since little experimental information about spin distribution is available. Thus the experimentally measured level density of this work does not correspond to the true multiplicity of states, and we use multiplicity $\Omega_l$ based on the experimental level density as:
\begin{equation}
\Omega_l(E) \propto \rho(E).
\label{eqn:5.2}
\end{equation}

For the present analysis, the microcanonical ensemble has been used since hot nuclei are better described statistically through such an ensemble~\cite{Morissay}. Within this framework, the nucleus is considered to be an isolated system with a well-defined energy. According to our definition of Eq.~(\ref{eqn:5.2}) the multiplicity of microstates $\Omega_l(E)$, which is obtained from the experimental level density $\rho(E)$, we define a "pseudo" entropy as:
\begin{align}
S(E) &=k_{B}\Omega_l(E) \nonumber \\
	 &=k_{B}\ln \frac {\rho(E)}{\rho_0} \nonumber \\
	&=k_{B} \ln \rho(E) + S_0,
\label{eqn:array2}
\end{align} 
where Boltzmann's constant $k_B$ is set to unity for the sake of simplicity. The normalization term $S_0$ is adjusted to fulfill the third law of thermodynamics, i.e., $S \rightarrow 0$ for $T \rightarrow 0$, $T$ being the temperature of the nucleus. Using the ground state band of the even-even $^{44}$Ti, the normalization term is found to be $S_0=-0.1$. Furthermore, the microcanonical temperature $T$ can be deduced from $S$ by
\begin{equation}
\frac{1}{T(E)} = \frac{\partial S}{\partial E}. \
\label{eqn:array3} 
\end{equation} 

\begin{figure}
\centering
\includegraphics[height=15cm]{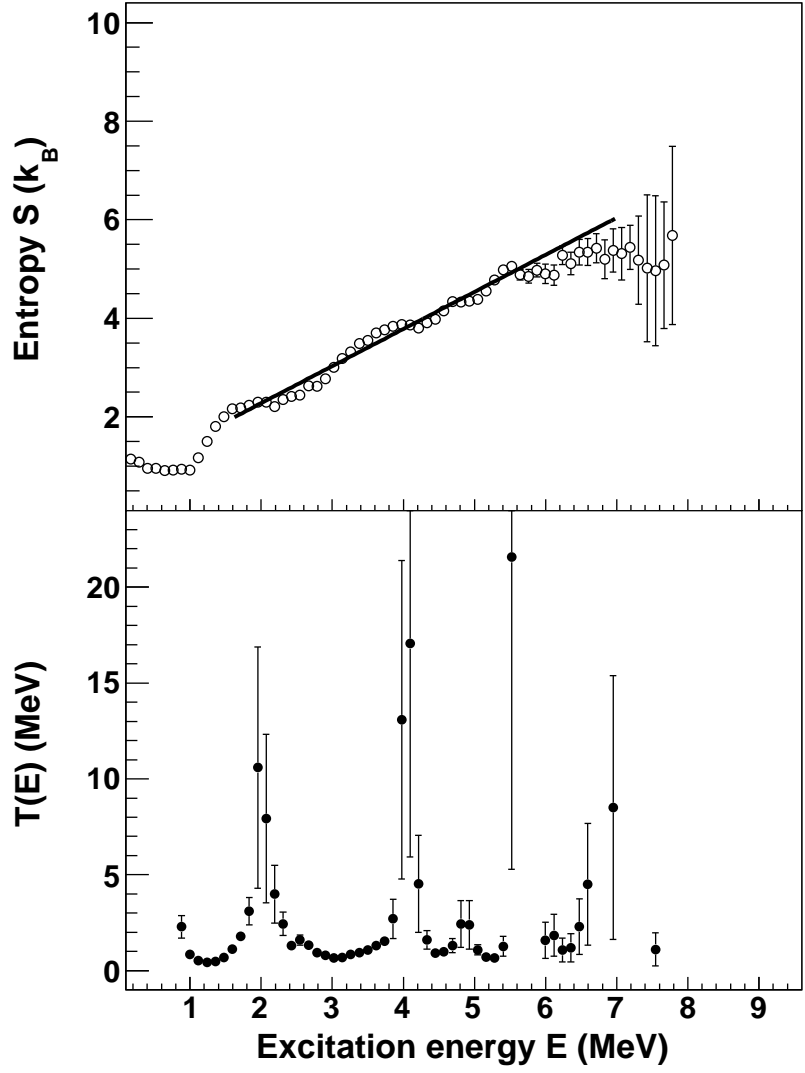}
\caption{Microcanonical entropy (upper pannel) and temperature (lower panel) as a function of excitation energy in $^{45}$Ti. A line is fitted (upper panel) for the data points lying between 1.5 and 7 MeV to determine an average temperature (see text).}
\label{fig:entropy1}
\end {figure}

Figure~\ref{fig:entropy1} shows the variation of entropy and temperature for $^{45}$Ti, within a microcanonical ensemble. The variations in entropy on a linear scale with the excitation energy are equivalent to the variations in the level density on a logarithmic scale. Generally, the most efficient way to generate additional states in an atomic nucleus is to break a nucleon Cooper pair from its core. The resulting two nucleons may thereby be thermally excited independently to the available single-particle levels around the Fermi surface. The entropy curve displays step like structures around 2 and 4 MeV excitation energies that can be interpreted as the breaking of one and two nucleon pairs, respectively.

From Fig.~\ref{fig:entropy1} the temperature fluctuations are seen to be more pronounced than for the entropy. This is because the fine structure of the entropy curve is enhanced in the temperature $T$ curve due to the differentiation. In spite of these fluctuations one can determine directly from the entropy an average temperature $\langle T \rangle$ for  $1.5 < E < 7.0$ MeV excitation energy region, as shown by a straight line in the Fig.~\ref{fig:entropy1} giving 1.4(1) MeV. The enhanced bump structures in the temperature spectra can be interpreted as the breaking of nucleon pairs. When particle pairs are broken, new degrees of freedom open up leading to an increase of $\rho(E)$ and decrease in temperature $T(E)$. In the temperature plot of Fig.~\ref{fig:entropy1} one can notice locations of negative slopes. The first such location appears at $E \sim$ 2.0 MeV which should be compared with twice the proton pairing gap parameter $2 \Delta_p$ of $^{45}$Ti, the minimum required energy to break a proton Cooper pair. Following the definition of~\cite{doba} one gets $2 \Delta_p = 1.784$ MeV, which is comparable to the location of the first temperature drop. The second location of a temperature drop occurs around $\sim 4.2$ MeV which can be interpreted as the onset of the four quasi-particle regime.

The experimental level density of $^{45}$Ti is compared with the BSFG level density as shown by Fig.~\ref{fig:nld1}. The figure shows that the BSFG model level density does not reproduce the detailed structures in the experimental level density. So, in order to investigate the level density further, a microscopic model has been applied.

\subsection{Combinatorial BCS Model of Nilsson Orbitals}

The model~\cite{Sc} is based on combining all the proton and neutron configurations within the Nilsson level scheme and using the concept of Bardeen-Cooper-Schrieffer (BCS) quasi-particles~\cite{BCS}. The single-particle energies $e_{\rm sp}$ are taken from the Nilsson model for an axially deformed core, where the deformation is described by the quadruple deformation parameter $\epsilon_2$. The quasi-particle excitation energies are described by
\begin{equation}
E_{\rm qp}(\Omega_\pi,\Omega_\nu) = \sum_{\Omega_\pi^\prime,\Omega_\nu^\prime} \left( e_{\rm qp}(\Omega_\pi^\prime) + e_{\rm qp}(\Omega_\nu^\prime) + V(\Omega_\pi^\prime,\Omega_\nu^\prime) \right) ,
\label{eqn:qpe}
\end{equation}
where $\Omega_\pi$ and $\Omega_\nu$ are the spin projections of protons and neutrons on to the symmetry axis, respectively, and $V$ is the residual interaction described by a Gaussian random distribution. The single quasi-particle energy $e_{\rm qp}$, characterized by the Fermi level $\lambda$ and pair-gap parameter $\Delta$, is defined as:
\begin{equation}
e_{\rm qp} = \sqrt{(e_{\rm sp}-\lambda)^2 + \Delta^2}. \
\label{eqn:qp}
\end{equation}
The total excitation energy is the sum of the quasi-particle energy $E_{\rm qp}(\Omega_\pi, \Omega_\nu)$, rotational excitations and vibrational excitations:
 \begin{equation}
E = E_{\rm qp}(\Omega_\pi, \Omega_\nu) + A_{\rm rot}R(R+1) + \hbar\omega_{\rm vib}\nu. \
\label{eqn:x}
\end{equation}
The rotational excitations are described by the rotational parameter $A_{\rm rot} = \hbar/2 {\cal I}$, with the moment of inertia ${\cal I}$ and the rotational quantum number $R$. The vibrational excitations are described by the phonon number $\nu = 0, 1, 2,\ldots$ and oscillator quantum energy $\hbar \omega_{\rm vib}$. 
At low excitation energy, the rotational parameter $A_{\rm rot}$ is for simplicity taken as the value deduced around the ground state $A_{\rm gs}$ of even-even nuclei in this mass region. For increasing excitation energy, the rotational parameter is decreased linearly according to
\begin{equation}
A_{\rm rot}(E) = A_{\rm gs} + \left(\frac{A_{\rm rigid}- A_{\rm gs}}{E_{\rm rigid}}\right)E,
\end{equation}
where we assume $A_{\rm rot}= A_{\rm rigid}$ for excitation energies above the excitation energy $ E_{\rm rigid}$. Since theoretical approaches seem to predict rigid body moments of inertia at the neutron separation energy, we set $ E_{\rm rigid}= S_n$. The rigid body value is calculated as
\begin{equation}
A_{\rm rigid} = \frac{5 \hbar}{4 M R_{A}^2 (1+0.31\epsilon_2)},
\end{equation}
where $M$ is the nuclear mass and $R_A$ is the nuclear radius. 

The spin ($I$) of each state is schematically calculated from the rotational quantum number ($R$) and the total projection ($K$) of the spin vector on the nuclear symmetry axis by
\begin{equation}
I(I+1) = R(R+1) + K^2.
\label{eqn:spin}
\end{equation}
The quantity $K$ is the sum of spin projections of protons and neutrons on the symmetry axis:
\begin{equation}
K=\sum_{{\Omega_{\pi}',\Omega_{\nu}'}} \Omega_{\pi}' + \Omega_{\nu}'.
\label{eqn:k}
\end{equation}

\begin{figure}
\centering
\includegraphics[height=15cm]{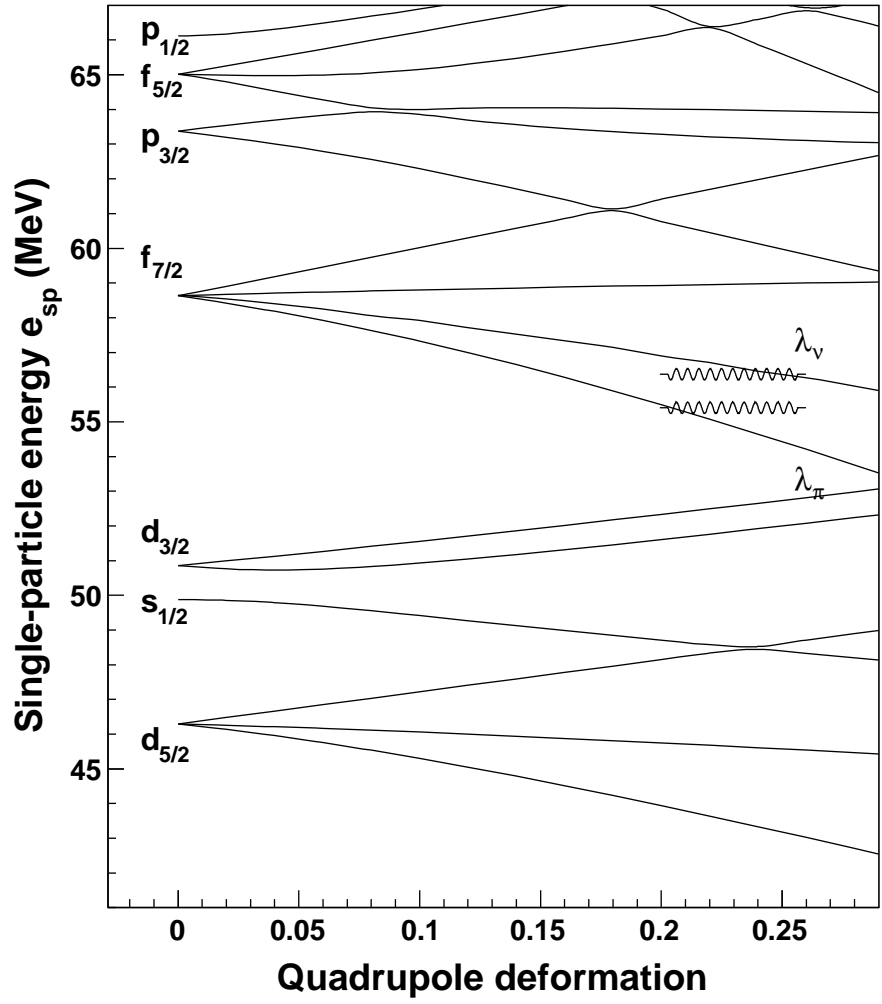}
\caption{Nilsson level scheme for $^{45}$Ti with single-particle energies as function of quadrupole deformation $\epsilon_2$.}
\label{fig:nilsson}
\end {figure}

The Nilsson single-particle orbitals calculated for $^{45}$Ti are drawn in Fig.~\ref{fig:nilsson}. The spin-orbit parameter $\kappa=0.066$ and the centrifugal parameter $\mu=0.32$ are taken from Ref.~\cite{white}. The main harmonic oscillator quantum energy is estimated by $\hbar \omega_0 = 1.2(41A^{-1/3})$ MeV. The vibrational quantum energy $\hbar \omega_{\rm vib} =$ 2.611 MeV is taken from the excitation energy of the first $0^+$ vibrational state in $^{46}$Ti. The Fermi levels for protons and neutrons are also shown in Fig.~\ref{fig:nilsson} for an estimated deformation of $\epsilon_2= 0.25$. Other parameters employed to calculate the level density of $^{45}$Ti are listed in Table~\ref{tab:2}.

\begin{figure}
\centering
\includegraphics[height=15cm]{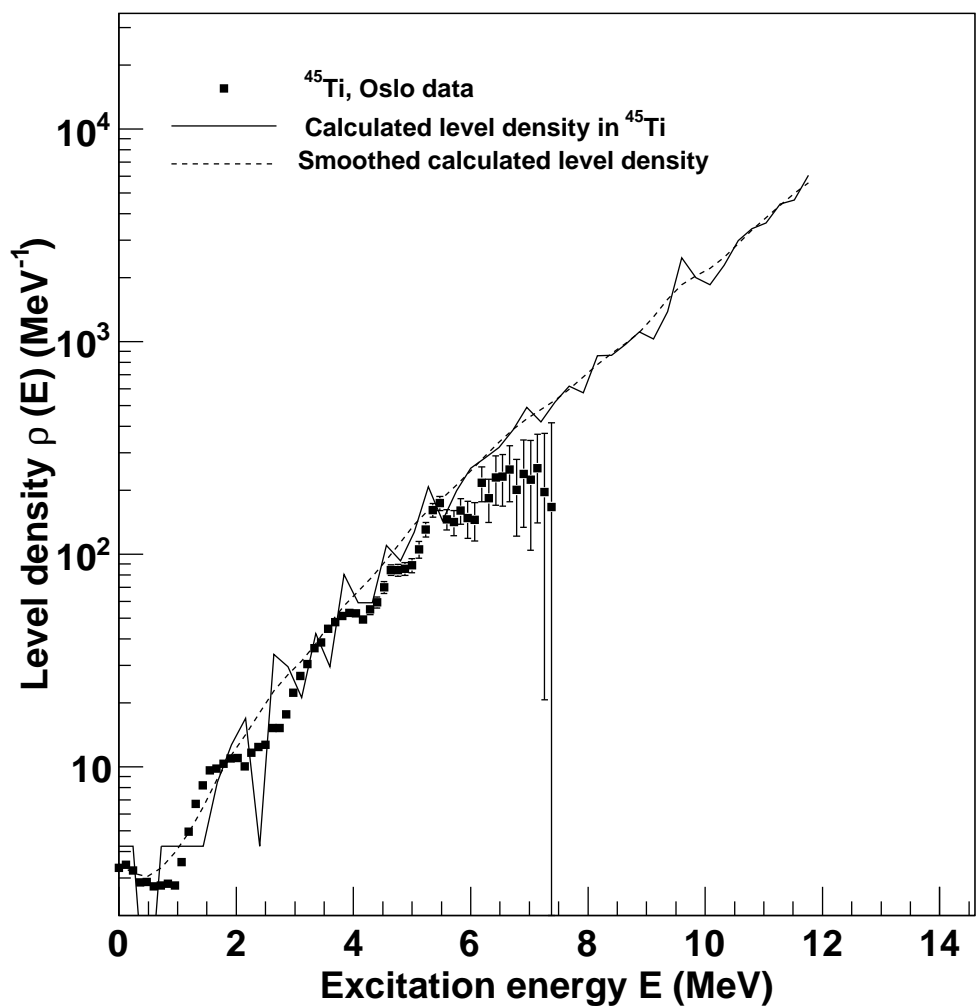}
\caption{Comparison between the experimental (filled squares) and calculated (solid line) level densities in $^{45}$Ti obtained with the combinatorial model~\cite{Sc}. The dashed curve represents the model prediction, but smoothed with 300 keV in order to match the experimental resolution.}
\label{fig:micro}
\end {figure}

\begin{figure}
\centering
\includegraphics[height=15cm]{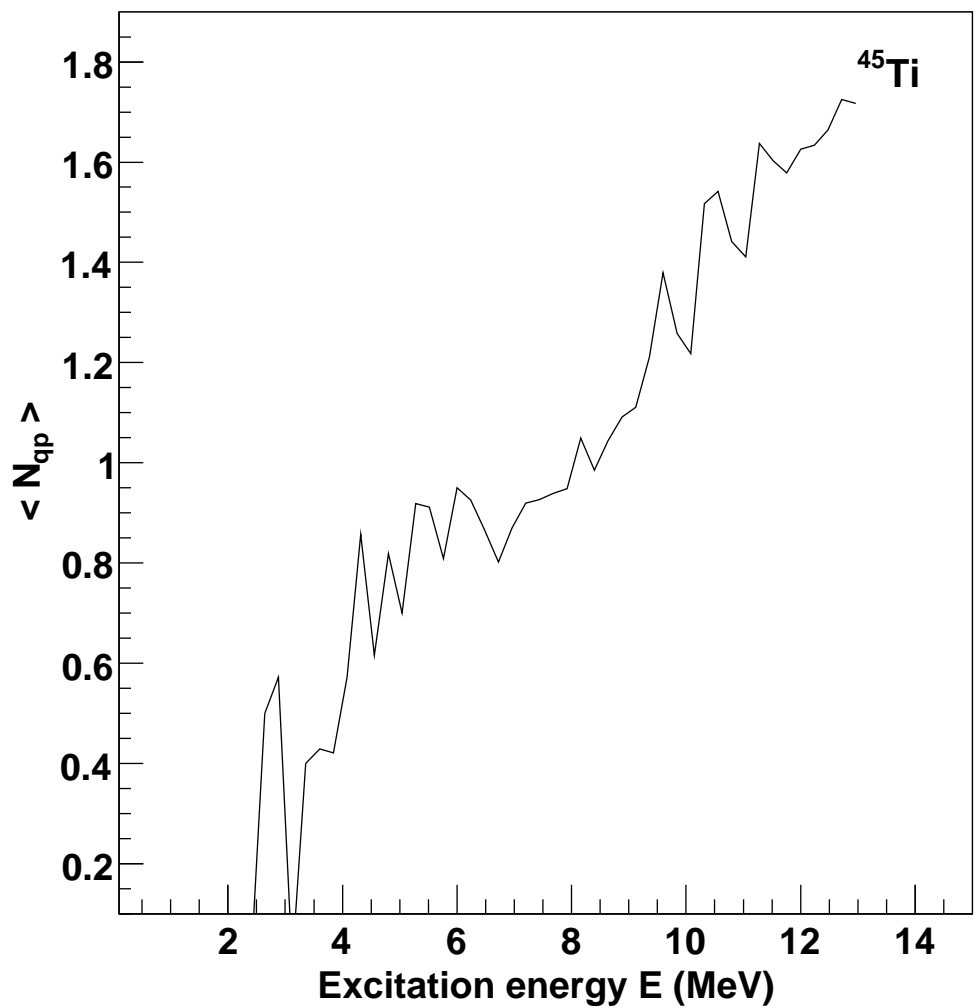}
\caption{Average number of broken Cooper-pairs in $^{45}$Ti calculated within the combinatorial model~\cite{Sc}.}
\label{fig:pairing}
\end {figure}

Figure~\ref{fig:micro} shows that the calculated level density for $^{45}$Ti describes satisfactorily the experimental level density. The general shape and the structural details of the level density are well reproduced. In Fig.~\ref{fig:micro} is also shown the calculated level density smoothed with the experimental resolution i.e., 300 keV. The smoothed curve agrees well with the experimental data points in absolute values, however, structural details are not accurately reproduced. Keeping in mind the simplicity of the model these results are very encouraging. The structures in the level density can be understood from Fig.~\ref{fig:pairing} where the average number of broken nucleon pairs $\langle N_{\rm qp} \rangle$ is plotted as a function of excitation energy. Here, the average number of pairs includes both proton and neutron pair breaking. Figure~\ref{fig:pairing} shows that the first pair breaks at $2\Delta \sim$ 2.5 MeV of excitation energy. The pair breaking process leads to an exponential increase of the level density, and it is also responsible for an overall increase in the level density at higher excitation energies. Rotational and vibrational excitations are less important. Even the shell gaps expected at $Z=N=20$ seem not to play a major role, probably because the gap between the $f_{7/2}$ and $d_{3/2}$ orbitals is reduced by the nuclear quadruple deformation. Indeed, Fig.~\ref{fig:nilsson} shows that the single-particle Nilsson orbitals are distributed rather uniformly in energy at $\epsilon_2=0.25$.

\begin{figure}
\centering
\includegraphics[height=15cm]{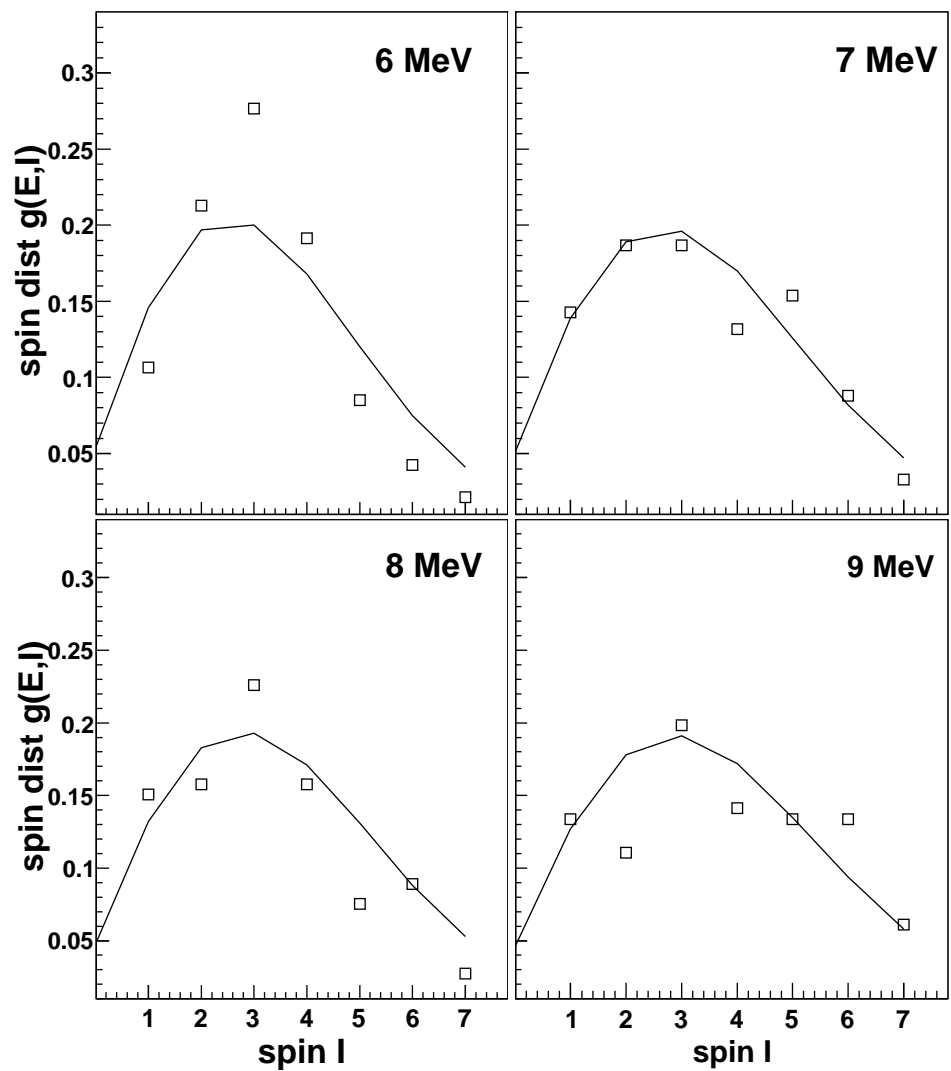}
\caption{Comparison of spin distributions in $^{45}$Ti determined by Eq.~(\ref{eqn:spin}) (open squares) with that determined by Eq.~(\ref{eqn:12}) (solid lines) at different excitation energy bins.}
\label{fig:spin}
\end {figure}

The spin distribution of our model can be compared with the Gilbert and Cameron spin distribution~\cite{GC}:
\begin{equation}
g(E,I) = \frac{2I+1}{2\sigma^2} \exp \left[ -(I+1/2)^2/2\sigma^2 \right]
\label{eqn:12}
\end{equation}
with the spin cut-off parameter $\sigma$ taken from Eq.~(\ref{eqn:8}). The two spin distributions (normalized to unity) are shown in Fig.~\ref{fig:spin}. The same parameters of $a$ and $E_1$ are used for the determination of $\sigma$, as those in the level density normalization. The comparison is made at four different excitation energy bins each having a width of 0.24 MeV. The general trend in the two distributions is surprisingly similar; however, some deviations are also apparent. This is mainly due to fluctuations originating from the low level density in this nucleus. The moment of inertia of  $^{45}$Ti is chosen to approach a rigid rotor at energies near and above $S_n$, which has been found appropriate theoretically in the medium mass region nuclei $A \sim 50 - 70$~\cite{alhassid}. The satisfactory resemblance of the two spin distributions indicates that our simplified treatment of determining the spin of levels through Eqs.~(\ref{eqn:spin}) and (\ref{eqn:k}) works well.

\begin{figure}
\centering
\includegraphics[height=15cm]{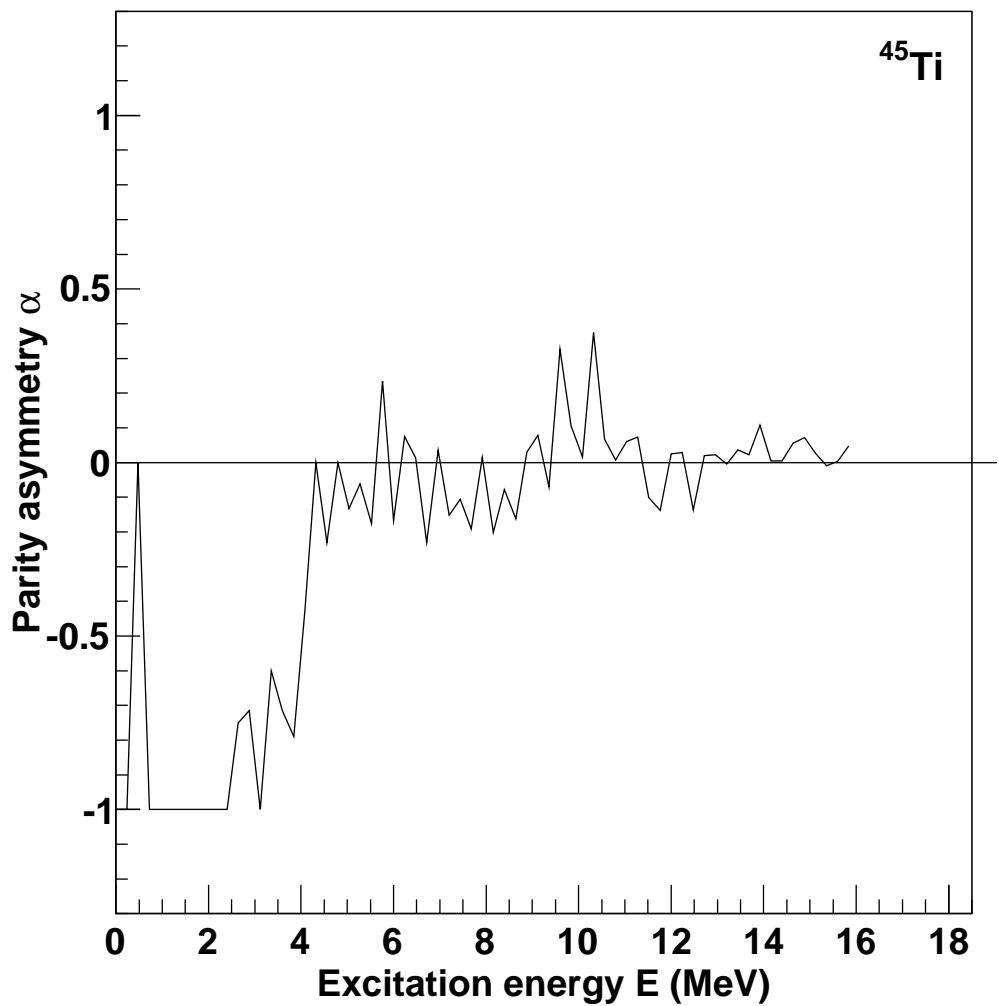}
\caption{Parity asymmetry in $^{45}$Ti as calculated in the combinatorial model~\cite{Sc}.}
\label{fig:parity}
\end {figure}

The parity distribution is a quantity that also reveals the presence (or absence) of shell gaps. In the extreme case, where only the $\pi f_{7/2}$ and $\nu f_{7/2}$ shells would be occupied by the valence nucleons, only negative parity states would appear. The parity asymmetry parameter $\alpha$ can be utilized to display the parity distribution in quasi-continuum and is defined by~\cite{parity}\begin{equation}
\alpha = \frac{\rho_+ - \rho_-}{\rho_+ + \rho_-},
\label{eqn:y}
\end{equation}
where $\rho_+$ and $\rho_-$ are the positive and negative parity level densities. An equal parity distribution would give $\rho_+ = \rho_-$, and thus $\alpha$ = 0. Other $\alpha$ values range from -1 to +1, i.e., from more negative parity states to more positive parity states. Figure~\ref{fig:parity} shows that there are more negative parity states ($\alpha < 0$) below 4 MeV, while at higher excitation energies where hole states in the $sd$ shell comes into play, the asymmetry is damped out giving an equal parity distribution. Thus, the parity calculations also confirm the absence of pronounced shell gaps in $^{45}$Ti.

\section{Gamma-ray strength function}

The $\gamma$-ray strength function can be defined as the distribution of the average decay probability as a function of $\gamma$-ray energy between levels in the quasi-continuum. The $\gamma$-ray strength function $f_{XL}$, where $X$ is the electromagnetic character and $L$ is the multipolarity, is related to the $\gamma$-ray transmission coefficient ${\cal T}_{XL}(E_\gamma)$ for multipole transitions of type $XL$ by 
\begin{equation}
{\cal T}(E_\gamma) = 2\pi\sum_{XL} E_\gamma^{2L+1}f_{XL}(E_\gamma).\
\label{eqn:9}
\end{equation}
According to the Weisskopf estimate~\cite{Blatt}, we assume that electric dipole $E1$ and magnetic dipole $M1$ transitions are dominant in a statistical nuclear decay. It is also assumed that the numbers of accessible positive and negative parity states are equal, i.e.,
\begin{equation}
\rho(E-E_\gamma, I_f, \pm\pi_f) = \frac {1}{2} \rho(E-E_\gamma, I_f). \
\label {eqn:10}
\end{equation}
The expression for the average total radiative width $\langle\Gamma_\gamma\rangle$~\cite{kopecky} for $s$-wave neutron resonances with spin $I_t \pm 1/2$ and parity $\pi_t$ at $E=S_n$, reduces to 
\begin{align}
\langle\Gamma_\gamma(S_n, I_t \pm 1/2)\rangle &= \frac{1}{4 \pi \rho(S_n, I_t \pm 1/2, \pi_t)} \int_0^{S_n}{\mathrm{d}}E_{\gamma} \nonumber\\
&  \times B{\mathcal{T}}(E_\gamma)\rho(S_n - E_\gamma)\nonumber\\
&  \times \sum_ {J=-1}^{1}{g(S_n-E_\gamma, I_t\pm1/2 + J)}.
\label{eqn:11}
\end{align}
 
As mentioned in Sect.~II, we extract the level density $\rho$ and transmission coefficient ${\cal T}$ from the primary $\gamma$-ray spectrum. The slope of ${\cal T}$ is determined by the transformation generator $\alpha$ (see Eq.(\ref{eqn:5})), which has already been determined during the normalization of $\rho$ in Sect.~II. However, the factor $B$ of Eq.~(\ref{eqn:5}) that determines the absolute value of the transmission coefficient, remains to be determined. The unknown factor $B$ can be determined using Eq.~(\ref{eqn:11}) if $\langle\Gamma_\gamma\rangle$ at $S_n$ is known from nucleon capture experiments. Unfortunately, the average total radiative width for $^{45}$Ti has not been measured. 
Therefore, we have applied $\langle\Gamma_\gamma\rangle = 1400(400)$ meV, which is the value of $^{47}$Ti from Ref.~\cite{Vonach} (see Table III). We deem this to be a reasonable approximation, since the level densities of odd-$A$ isotopes $^{47,49}$Ti are comparable at $S_n$ (see Fig. 1). In addition, $^{45,47}$Ti are not closed shell nuclei, so that abrupt changes in the nuclear structure are not expected. Therefore, similar decay properties at $S_n$ might be a valid assumption.

From our calculation in Fig.~\ref{fig:parity} it is obvious that the assumption of equal parity, as described by Eq.~(\ref{eqn:10}), does not hold below 4 MeV of excitation energy. This means that more $M1$ transitions are expected for $\gamma$-decays from $S_n = 9.53$ MeV to levels below 4 MeV than above 4 MeV where $E1$ transitions are prominent. Thus, it should be pointed out that in the evaluation of $B$ this enhanced $M1$ transitions over $E1$-transitions below 4 MeV might affect the absolute normalization of the $\gamma$-ray strength function.

\begin{figure}

\includegraphics[height=15cm]{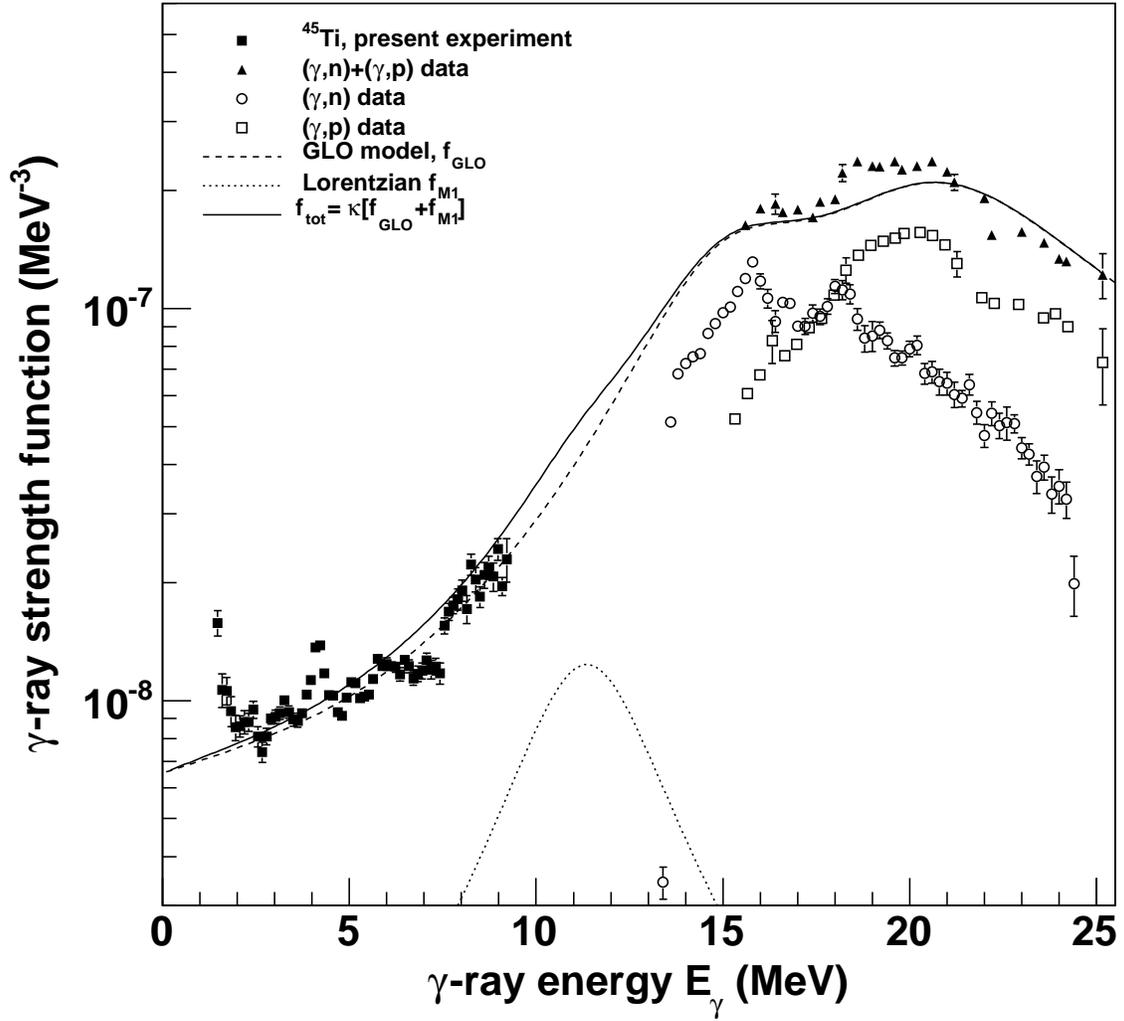}
\caption{Normalized $\gamma$-ray strength function for $^{45}$Ti as a function of $\gamma$-ray energy. The Oslo data (filled squares) are compared with the GLO model~\cite{6.15} (solid line). In addition, the strength function of $^{46}$Ti obtained from the data of ($\gamma$,n) reaction~\cite{gamma_n} (open circles), ($\gamma$,p) reaction~\cite{gamma_p} (open squares), and their sum (filled triangles) have also been drawn at $\gamma$-ray energies above the particle threshold.}
\label{fig:rsf}
\end {figure}

The resulting normalized $\gamma$-ray strength function of $^{45}$Ti is shown in Fig.~\ref{fig:rsf}. In the same figure the experimental data of $^{45}$Ti and the Generalized Lorentzian Model (GLO)~\cite{RIPL} are compared. The model is used to describe the giant electric dipole resonance (GEDR) at low $\gamma$ energies and at resonance energies. In the lower energy region this model gives a non-zero finite value of the dipole strength function in the limit of $E_\gamma \rightarrow 0$. The GLO model as proposed by Kopecky and Chrien~\cite{6.15}, describes the strength function as:
\begin{align}
f_{GLO} &= \frac {1}{3 \pi^2 \hbar^2 c^2} \sigma_{E1} \Gamma_{E1} \nonumber \\
& \times [ E_\gamma \frac{\Gamma_k(E_\gamma,T)}{(E_\gamma^2-E_{E1}^2)^2 + E_\gamma^2\Gamma_k^2(E_\gamma,T)} \nonumber \\
& + 0.7 \frac{\Gamma_k(E_\gamma = 0,T)}{E_{E1}^3}], 
\label{eqn:16}
\end{align}
where $\sigma_{E1}$, $\Gamma_{E1}$ and $E_{E1}$ are the cross-section, width and energy of the centroid of the GEDR, respectively. The energy and temperature-dependent width $\Gamma_k$ is given by~\cite{RIPL1}
\begin{equation}
\Gamma_k(E_\gamma,T) = \frac{\Gamma_{E1}}{E_{E1}^2}(E_\gamma^2 + 4 \pi^2 T^2).
\label{eqn:17}
\end{equation} 

The even-mass titanium isotopes have ground state nuclear deformation~\cite{RIPL} indicating that the GEDR resonance contains two components. In Table~\ref{tab:3} the two sets of GEDR parameters for the ground state deformation $\epsilon_2 \sim 0.25$ (interpolated between the known neighboring nuclei) are listed using the systematics of Ref.~\cite{RIPL}. Furthermore, we assume that the $\gamma$-ray strength function is independent of excitation energy, i.e., we use a constant temperature $T= 1.4$ MeV. This constant temperature approach is adopted in order to be consistent with the Brink-Axel hypothesis in Sect.~II, where the transmission coefficient ${\cal T}(E_\gamma)$ is assumed to be temperature independent.

The magnetic dipole $M1$ radiation, supposed to be governed by the giant magnetic dipole (GMDR) spin-flip $M1$ resonance radiation~\cite{voinov}, is described by a Lorentzian~\cite{GMDR}
\begin{equation}
f_{M1}(E_\gamma) = \frac{1}{3\pi^2 h^2 c^2} \frac{\sigma_{M1}E_\gamma \Gamma_{M1}^2}{(E_\gamma^2 - E_{M1}^2)^2 + E_\gamma^2 \Gamma_{M1}^2},
\label{eqn:18}
\end{equation}
where $\sigma_{M1}$, $\Gamma_{M1}$, and $E_{M1}$ are the GMDR parameters deduced from systematics given in Ref.~\cite{RIPL}.

The total model $\gamma$-ray strength function is given by
\begin{equation}
f_{\mathrm{tot}} = \kappa \left[ f_{E1,1} + f_{E1,2} + f_{M1} \right],
\label {eqn:19}
\end{equation}
where the factor $\kappa$ is used to scale the model strength function to fit the experimental values. The value of $\kappa$ is expected to deviate from unity, since only approximate values of the average resonance spacings $D$ and the total average radiative width data $\langle\Gamma\rangle$ have been used for the absolute normalization of the strength function. 
 
In order to increase the confidence of our normalization, the strength functions deduced from photoneutron cross-section data~\cite{gamma_n} of the $^{46}$Ti($\gamma$,n) reaction and from photoproton cross-section data~\cite{gamma_p} of the $^{46}$Ti($\gamma$,p) reaction are displayed together in Fig.~\ref{fig:rsf}. In order to transform cross-section into $\gamma$-ray strength function the following relation~\cite{RIPL} is employed:
\begin{equation}
f(E_\gamma)= \frac {1} {3\pi^2 \hbar^2 c^2} \frac {\sigma(E_\gamma)}{E_\gamma}. \
\label{eq:13}
\end{equation}
The data of the two photoabsorption experiments around $\gamma$-ray energies $E_\gamma = 15 - 26$ MeV are shown as filled triangles in Fig.~\ref{fig:rsf}. By assuming that the $\gamma$-ray strength function is a weak function of nuclear mass $A$, we can expect that $\gamma$-ray strength functions for neighboring nuclei are comparable. From Fig.~\ref{fig:rsf}, one finds that the increase of the $^{45}$Ti $\gamma$-ray strength function with $\gamma$-ray energy is well in line with the ($\gamma$,n) + ($\gamma$,p) data. Note that the $^{46}$Ti($\gamma$,p) cross section below $E_\gamma$ < 15 MeV approaches the particle threshold energies and are less reliable.
  
The shape of the GEDR cross section obtained from the sum of ($\gamma$,n) and ($\gamma$,p) data~\cite{gamma_n,gamma_p} clearly shows a splitting of the GEDR into two or more resonances. This splitting is a signature of the ground state deformation of the nucleus, which justifies our treatment of the GEDR, i.e., fitting two distributions using two sets of resonance parameters. 

Figure~\ref{fig:rsf} shows that the Oslo data are well described by the GLO model in the energy region $\sim 2.5 -9$ MeV. However, at low $\gamma$-ray energies ($E_\gamma < 2.5$ MeV) an enhancement in the $\gamma$-ray strength function compared to the GLO model has been observed. This upbend has been observed in several nuclei with mass number $A < 100$ (see e.g.,~Ref.~\cite{Sc} and references therein). The physical origin of this enhancement has not been fully understood yet, and no theoretical model accounts for such a behavior of the nucleus at low $\gamma$-ray energies.

\section{Conclusions} 

The level density and $\gamma$-ray strength function of $^{45}$Ti have been measured using the $^{46}$Ti$(p,d)^{45}$Ti reaction and the Oslo method. The thermodynamic quantities entropy and temperature of the microcanonical ensemble are extracted from the measured level densities. The average temperature for $^{45}$Ti is found to be 1.4 MeV. The experimental level density is also compared with a combinatorial BCS model using Nilsson orbitals. This model describes satisfactorily the general increase and structural details of the experimental level density.

The generalized Lorentzian model (GLO) has been compared with the experimental $\gamma$-ray strength function. The GLO model describes the Oslo data well in the energy region $E_\gamma \sim 2.5 - 9$ MeV. However, at $E_\gamma < 2.5$ MeV an enhancement in the $\gamma$-ray strength function compared to the GLO model has been observed. This is very interesting since a similar $\gamma$-decay behavior has been observed in several other light mass nuclei, which is not accounted for yet by present theories.

\onecolumngrid

\begin{table}
\caption{Parameters used for the back-shifted Fermi gas level density.}

\begin{tabular}{cccccccc}
\hline
 $S_n$ & $a$ &		$E_1$ &	$\sigma$ &	$\rho(S_n)$ &$\eta$	\\ 
 (MeV) & (MeV$^{-1}$) & (MeV) &  & (10$^3$ MeV$^{-1}$) & \\
\hline\hline

9.53 & 5.62 & -0.90 & 3.48 & 1.4(7)$^{a}$ & 0.5\\
\hline\hline

\end{tabular} 

$^{a}$ Estimated from the systematics of Fig.~\ref{fig:estimation} (see text). \\
\label{tab:tab1}
\end{table}

\begin{table}
\caption{Parameters used in combinatorial BCS model for the calculation of level density.}
\begin{tabular}{cccccccc}
\hline
$\epsilon_2$  & $\Delta_\pi$ & $\Delta_\nu$ & $\kappa$ & $\mu$ & $A_{\rm gs}$ & $\lambda_\pi$ & $\lambda_\nu$ \\
	& (MeV) & (MeV) &	&	&	(MeV)	&	(MeV)	&	(MeV)	\\
\hline\hline

0.25 & 0.892 & 1.350 & 0.066 & 0.32 & 0.075 & 54.913 & 56.872 \\  
\hline\hline
\end{tabular}
\label{tab:2}
\end{table}

\begin{table}
\caption{GEDR and GMDR parameters determined using the systematics given in~\cite{RIPL}.}
\begin{tabular}{ccc|ccc|ccc|c|c}
\hline
$E_{E1,1}$ & $\Gamma_{E1,1}$ & $\sigma_{E1,1}$ & $E_{E1,2}$ & $\Gamma_{E1,2}$ & $\sigma_{E1,2}$ & $E_{M1}$ & $\Gamma_{M1}$ & $\sigma_{M1}$ & $\langle \Gamma_\gamma \rangle$ & $\kappa$ \\ 
(MeV) & (MeV) & (mb) & (MeV) & (MeV) & (mb) & (MeV) & (MeV) & (mb) &(meV) & \\
\hline\hline
16.20 & 5.31 & 22.26 & 22.52 & 9.96 & 44.53 & 11.53 & 4.0 & 1.23 & 1400(400)$^{b}$ & 1.05\\
\hline\hline 
\end{tabular}

$^{b}$ The $s$-wave average total radiative width $\langle \Gamma_\gamma \rangle$ of $^{47}$Ti~\cite{Vonach} has been used for normalization.\\
\label{tab:3}

\end{table}

\newpage

\newpage

\newpage

\newpage

\newpage

\newpage

\newpage

\newpage

\newpage

\newpage

\end{document}